\documentclass[11pt]{article}
\usepackage[utf8]{inputenc}
\usepackage{lmodern}
\usepackage[T1]{fontenc}
\usepackage[top=1in, bottom=1.in, left=1in, right=1in]{geometry}
\usepackage{graphicx}
\usepackage{longtable}
\usepackage{float}
\usepackage{wrapfig}
\usepackage{rotating}
\usepackage[normalem]{ulem}
\usepackage{amsmath}
\usepackage{textcomp}
\usepackage{marvosym}
\usepackage{wasysym}
\usepackage{amssymb}
\usepackage{amsmath}
\usepackage[theorems, skins]{tcolorbox}
\usepackage[version=3]{mhchem}
\usepackage[numbers,super,sort&compress]{natbib}
\usepackage{natmove}
\usepackage{url}

\usepackage[frozencache=true,cachedir=minted-cache]{minted}

\usepackage[strings]{underscore}
\usepackage[linktocpage,pdfstartview=FitH,colorlinks,
linkcolor=blue,anchorcolor=blue,
citecolor=blue,filecolor=blue,menucolor=blue,urlcolor=blue]{hyperref}
\usepackage{attachfile}
\usepackage{setspace}
\usepackage{tcolorbox}
\usepackage{xcolor} 
\definecolor{lightbrown}{RGB}{210,180,140}
\tcbuselibrary{listingsutf8}  
\tcbset{colback=lightbrown!20, colframe=black, boxrule=0.5pt, arc=3pt, boxsep=5pt}
\definecolor{lightgray}{RGB}{230,230,230}
\tcbset{colframe=black, boxrule=0.5pt, arc=3pt, boxsep=5pt}
\setminted{frame=none}  
\author{John Kitchin}
\date{}
\title{The Evolving Role of Programming and LLMs in the Development of Self-Driving Laboratories}
\begin{document}

\author{John Kitchin\thanks{Corresponding author: jkitchin@andrew.cmu.edu} \\
Department of Chemical Engineering\\
Carnegie Mellon University,\\
5000 Forbes Ave, Pittsburgh, PA 15213, USA}

\maketitle

\begin{abstract}
Machine learning and automation are transforming scientific research, yet the implementation of self-driving laboratories (SDLs) remains costly and complex, and it remains difficult to learn how to use these facilities. To address this, we introduce Claude-Light, a lightweight, remotely accessible instrument designed for prototyping automation algorithms and machine learning workflows. Claude-Light integrates a REST API, a Raspberry Pi-based control system, and an RGB LED with a photometer that measures ten spectral outputs, providing a controlled but realistic experimental environment. This device enables users to explore automation at multiple levels, from basic programming and experimental design to machine learning-driven optimization. We demonstrate the application of Claude-Light in structured automation approaches, including traditional scripting, statistical design of experiments, and active learning methods. Additionally, we explore the role of large language models (LLMs) in laboratory automation, highlighting their use in instrument selection, structured data extraction, function calling, and code generation. While LLMs present new opportunities for streamlining automation, they also introduce challenges related to reproducibility, security, and reliability. We discuss strategies to mitigate these risks while leveraging LLMs for enhanced efficiency in self-driving laboratories. Claude-Light provides a practical and accessible platform for students and researchers to develop automation skills and test algorithms before deploying them in larger-scale SDLs. By lowering the barrier to entry for automation in scientific research, this tool facilitates broader adoption of AI-driven experimentation and fosters innovation in autonomous laboratories.
\end{abstract}
\section{Introduction}
\label{sec:orga0e4ec8}

Machine learning in science has grown with significant speed in the past five years. We are now at a stage where often the algorithms are limited by data, and many efforts are underway to accelerate the rate at which data can be acquired. One approach often suggested is the self-driving laboratory \cite{haese-2019-next-gener,hysmith-2024-futur-self,tom-2024-self-drivin,lo-2024-review-low} which leverages automation with closed-loop machine learning algorithms to efficiently and quickly acquire the data required. Several reports suggest these approaches have been used to discover new materials \cite{szymanski-2023-auton-labor} along with publications on the software behind the automated lab  \cite{fei-2024-alabos}. Some argue full automation is not the solution, and that there are important reasons to keep humans in the loop  \cite{scheurer-2025-role-human}. There is consensus either way that automation will play an important role in advancing science and machine learning.

Applications of automation and machine learning are very common in the literature, but they are still not the way most science is done. It remains challenging to teach students how to do it. Automated labs and experiments are expensive to create and operate, and require a lot of different skills that take a long time to master in domain knowledge, programming and data science. Thus, there is a need for smaller scale systems to learn and develop methods on. An early example of a system like this was SDL-Light, a small, remotely accessible instrument \cite{baird-2022-what-is,baird-2023-build-hello} that could be automated for experimentation and analysis. Briefly, this instrument had an RGB LED with three settable channels for each color. A photometer sensor then could detect 10 channels of color intensity. Users could control the instrument remotely from a Jupyter notebook to practice using machine learning algorithms to perform optimizations or other experiments.

In this paper we introduce a variation of SDL-Light that we call Claude-Light (\url{https://github.com/jkitchin/claude-light}) as a tool for prototyping automation algorithms. We use this instrument to illustrate a hierarchy of automation concepts. We first use conventional programming tools to build up to using machine learning algorithms to build models that map the input setting to output measurements.

Large language models are one of the newest approaches to using machine learning in automation and self-driving laboratories. Although this approach is also just "code", we find the paradigm of using LLMs in code to be so different from conventional code that it is worth introducing these new concepts by example with Claude-Light.

This paper is written differently than most papers and it includes a lot of code examples. This was a choice made to provide concrete examples of what it means to use code in these problems. The intended audience for this work is scientists and engineers with basic programming skills who are interested in how automation and LLMs could influence their work. We aim to provide a framework to think about the advantages and disadvantages of each approach.
\section{Automating a remotely accessible instrument}
\label{sec:org0bc6864}

Our device builds on SDL-Light \cite{baird-2022-what-is,baird-2023-build-hello} with the following adaptions. We use a REST API instead of MQTT because we are more familiar with it and it was easier to develop. Specifically, by using flask we can provide a variety of interfaces at different levels of sophistication including a browser interface and API access. We use a Raspberry Pi (Rpi) instead of a PicoW because the Rpi has a built in Ethernet port, and it is easier to debug since we connect to it by secure shell, and it has a display port. We use float inputs from 0-1 instead of integers. With the Rpi we can use the full Python language which we are more familiar with than MicroPython which is used on the PicoW, and it has more functionality. We also integrated a camera into part of the interface so you can see what color the LED looks like in a measurement, especially when it is truly remote.

\begin{figure}[H]
\centering
\includegraphics[width=3in]{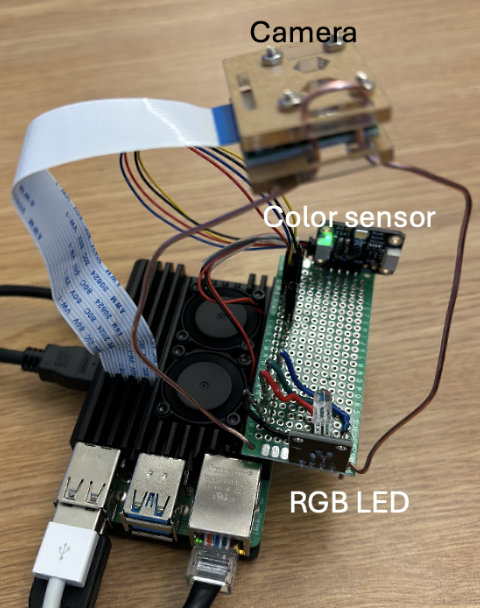}
\caption{\label{claude-light}Labeled image of Claude-Light.}
\end{figure}

Details of the instrument can be found online \cite{claude-light}.  A short summary is there are three inputs one can set: the R, G and B levels of an RGB LED. There is a photometer that measures 10 outputs: 8 different wavelengths, near IR and a clear channel of light. An image of Claude-Light is shown in Figure \ref{claude-light}. It is not the most important part of the paper, and we do not discuss it further here.

\begin{figure}[htbp]
\centering
\includegraphics[width=.9\linewidth]{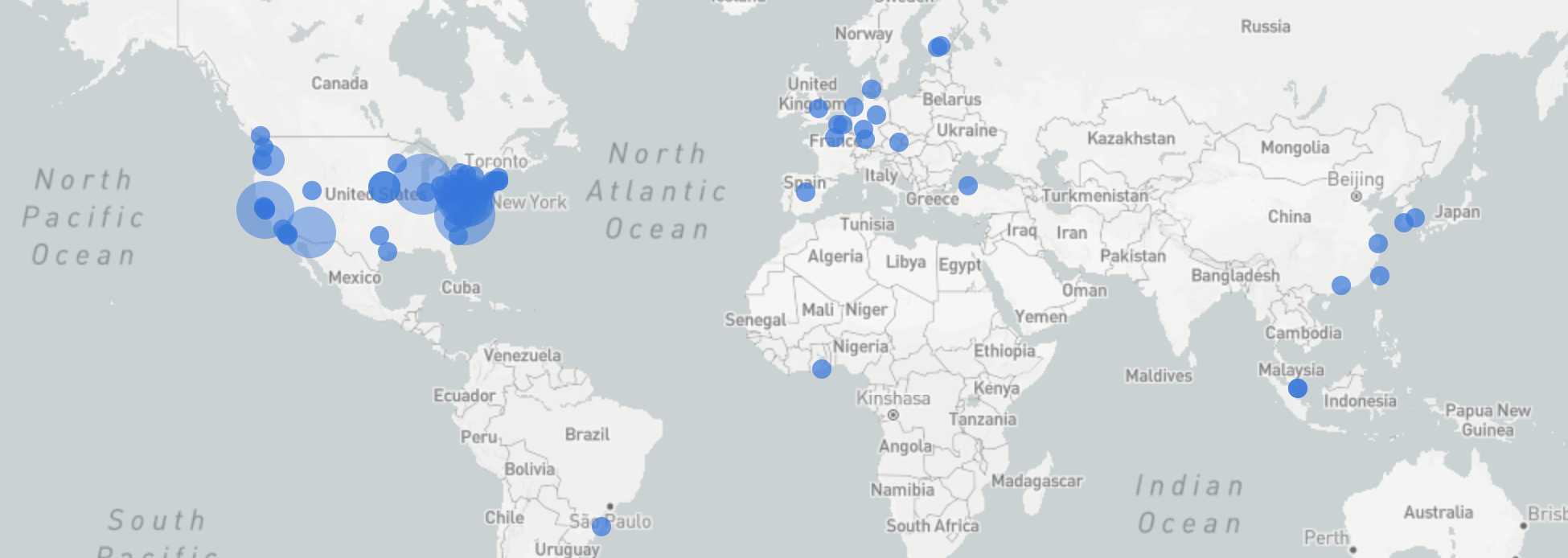}
\caption{\label{world-usage}Usage of Claude-Light around the world. The size of each dot is proportional to the number of users from that area. Last accessed February 20, 2025.}
\end{figure}

There are a number of features that the device has that make it useful for prototyping automation programs and algorithms. First, it is a real process with some noise in the measurements, reflecting real measurements. The outputs are practically linear in the inputs, so the modeling is not very difficult. That enables one to focus on the automation and design of experiments instead of the modeling. The photometer is open to the environment, and background illumination from daylight and office lights affect the measurements, including saturating some color channels at some times of the day. A measurement typically takes 1-2 seconds. We have used the device in classes with \textasciitilde{}50 students using it at a time with limited issues. The device has been used by 1800+ unique IP addresses around the world (Fig. \ref{world-usage}) and it has performed 73K+ experiments since around August of 2024.

\subsection{Basic REST API}
\label{sec:org4448841}

The Raspberry Pi acts as a web server to provide REST API to the instrument through a Flask app. This provides several endpoints for use in a browser, curl or any language that supports HTTP requests. The browser has access to the following endpoints that render as interactive forms that return data and an image of the device. These are not the main focus in this work, but we mention them to illustrate there are many ways to interact with a remote instrument.

\begin{description}
\item[{/gm}] the Green Machine that has a single input for the green input, and the intensity at 515nm for the output.
\item[{/rgb}] the RGB Machine with three inputs for the R, G and B levels, and all 10 outputs in tabular form
\end{description}

There is also an endpoint for programs to use:

\begin{description}
\item[{/api}] R, G, B in, and all the results out in json.
\end{description}

To illustrate flexibility we show two ways to interact with the instrument by the REST API. In the first example we use \texttt{curl} and pipe the output through \texttt{jq} to extract the result at 515nm, which is an approximately green color. \texttt{curl} is a command line utility to make http requests, and \texttt{jq} is a command line tool to process json data. The output here is an integer for the intensity at 515nm, which has a 16-bit range.

\begin{tcolorbox}[colback=lightbrown!20, colframe=black, boxrule=0.5pt]
\begin{minted}[frame=none,fontsize=\scriptsize,linenos=]{sh}
curl -s "https://claude-light.cheme.cmu.edu/api?R=0.12&G=0.45&B=1" | jq -M '.out."515nm"'
\end{minted}
\end{tcolorbox}
\begin{tcolorbox}[colback=lightgray!15]
\scriptsize \phantomsection
\label{}
\begin{verbatim}
29270
\end{verbatim}

\end{tcolorbox}
Here we illustrate the use of the requests library in Python, where we again extract the output intensity at 515nm.

\begin{tcolorbox}[colback=lightbrown!20, colframe=black, boxrule=0.5pt]
\begin{minted}[frame=none,fontsize=\scriptsize,linenos=]{python}
import requests
resp = requests.get('https://claude-light.cheme.cmu.edu/api',
                    params={'R': 0.12, 'G':0.45, 'B': 1})
print(resp.json()['out']['515nm'])
\end{minted}
\end{tcolorbox}
\phantomsection
\label{}
\begin{tcolorbox}[colback=lightgray!15]
\scriptsize \begin{verbatim}
31676
\end{verbatim}

\end{tcolorbox}

These are illustrative examples, and can be reproduced in any computer language that supports HTTP requests (these examples are GET requests). This flexibility is valuable because it allows many users to integrate the instrument into their work environment, and provides alternative options for debugging.
\subsection{Python API}
\label{sec:orgde5d7f3}

Shell scripting (using curl) is not very expressive, or flexible. Even the Python example above would lead to lengthy code to do anything of interest. Claude-Light has a Python library \cite{claude-light-python} that provides convenient access to four different instruments:

\begin{description}
\item[{GreenMachine1}] one input, one output
\item[{GreenMachine3}] one input, three outputs
\item[{CLRGB}] three inputs, three outputs
\item[{CLLight}] three inputs, 10 outputs
\end{description}

These are all classes with a \texttt{\_\_call\_\_} method defined, so we make an instance of the class and then call it with arguments to "make measurements". Here is an example usage. Note that it takes only three lines of code to make a measurement.

\begin{tcolorbox}[colback=lightbrown!20, colframe=black, boxrule=0.5pt]
\begin{minted}[frame=none,fontsize=\scriptsize,linenos=]{python}
from claude_light import CLRGB
cl = CLRGB()
print(cl(G=0.5))
\end{minted}
\end{tcolorbox}
\phantomsection
\label{}
\begin{tcolorbox}[colback=lightgray!15]
\scriptsize \begin{verbatim}
[5995, 34212, 1663]
\end{verbatim}

\end{tcolorbox}
\subsubsection{API in a loop}
\label{sec:org24c01a5}

Once we have the API it becomes easy to wrap it in a loop. Here we plot the output at 515nm as a function of the green channel input with a line fitted to it. This is the most basic design of experiment with evenly spaced points and a regression analysis. This already shows the first stage of automation, where the experiment is "automatically" run inside the loop. It is evident substantial compression of syntax has taken place, the majority of this code is making a figure, and it is a single line where the loop occurs to make the measurements.

\begin{tcolorbox}[colback=lightbrown!20, colframe=black, boxrule=0.5pt]
\begin{minted}[frame=none,fontsize=\scriptsize,linenos=]{python}
import numpy as np
import matplotlib.pyplot as plt

g = np.linspace(0, 1, 5)
G = [cl(G=x)[1] for x in g]

p = np.polyfit(g, G, 1)  # fits a line: G = m g + b

print(f'The slope is {p[0]:1.2f}, intercept is {p[1]:1.2f}')

plt.plot(g, G, 'go', label='data')
plt.plot(g, np.polyval(p, g), 'g-', label='fit')
plt.xlabel('G LED input level')
plt.ylabel('intensity at 515nm');
\end{minted}
\end{tcolorbox}
\phantomsection
\label{}
\begin{tcolorbox}[colback=lightgray!15]
\scriptsize \begin{verbatim}
The slope is 62466.40, intercept is 3110.20
\end{verbatim}

\end{tcolorbox}
\begin{center}
\includegraphics[width=.9\linewidth]{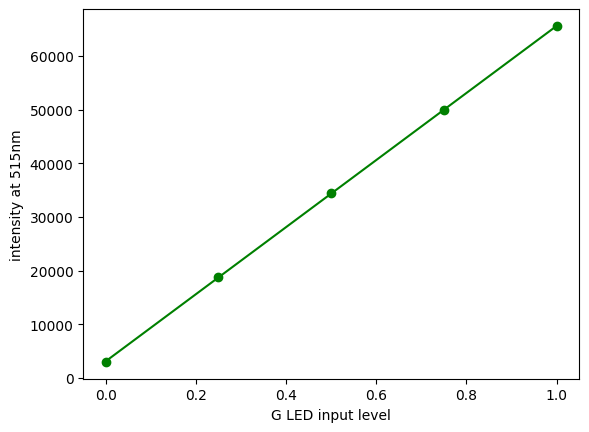}
\end{center}

It is straightforward to extend this to multivariate inputs, as well as multiple outputs. That data can be fitted with linear models conveniently. We note here that as presented, the data is ephemeral; it lives in memory until the program closes, and then it only exists in graphical form. It is straightforward to save the data to disk in a broad range of formats.
\subsubsection{API + design of experiments}
\label{sec:orgfdbe750}

It is a small step to transition from a loop over one variable to a design of experiment. Here we use a Latin Square design to test effect sizes of the inputs on the outputs. There are three inputs, the R, G and B LED settings. We want to know their effect size on the outputs of the CLRGB instrument (there are three outputs for 630nm, 515nm, and 445nm, which are red, greenish and bluish). A latin square design is an efficient way to vary all three inputs simultaneously, and in only nine experiments estimate the effect sizes  \cite{box-1978-statis-exper}. We leverage the pycse package \cite{pycse} written by the author to generate the design, and then we loop through each experiment, accumulate the results and do the analysis. Many other designs could be considered including surface response models. In this example, we only consider the effect of the Red input channel on the 630nm (red) output channel.

\begin{tcolorbox}[colback=lightbrown!20, colframe=black, boxrule=0.5pt]
\begin{minted}[frame=none,fontsize=\scriptsize,linenos=]{python}
from claude_light import CLRGB
from pycse.sklearn.lhc import LatinSquare
import pandas as pd

ls = LatinSquare({'R': [0, 0.5, 1.0],
                  'G': [0, 0.5, 1.0],
                  'B': [0, 0.5, 1.0]})

design = ls.design()

rgb = CLRGB()
results = pd.DataFrame([rgb(*row) for i, row in
                        design.iterrows()],
                       columns=['Ro', 'Go', 'Bo'])

ls.fit(design, results['Ro'])
print(ls.anova())
\end{minted}
\end{tcolorbox}
\phantomsection
\label{}
\begin{tcolorbox}[colback=lightgray!15]
\scriptsize \begin{verbatim}
Ro  effect      F-score (fc=19.0) Significant
0   R_effect    2139662.972784        True
1   G_effect       4417.979772        True
2   B_effect       1228.199706        True
3  residuals               1.0       False
\end{verbatim}

\end{tcolorbox}

It is not surprising to find out that the Red channel has the biggest effect on the Red output. It is a little surprising that there is some effect of the red input on the green and blue outputs. That is probably because the RGB channels are not pure, and have a distribution of wavelengths they output, with overlap on these output channels. That is not the point here though, the main point is that the API enables easy automation, here that also happens in a loop that iterates over each experiment and aggregates them for analysis.
\subsubsection{API + ML}
\label{sec:org1a01156}

Today it is common to use machine learning models instead of the fixed mathematical forms of conventional design of experiments. ML models accommodate more nonlinearity, usually at the expense of data requirements. In this example, we use a Gaussian Process model with a linear kernel. We specifically choose this kernel because we have prior knowledge that the outputs are basically linear in the inputs. A Relu neural network would give similar results here. A more nonlinear activation function or kernel would require a lot more data than the twenty experiments we do here. Previous experience suggested it was useful to scale the outputs because the inputs vary from 0 to 1, while the outputs vary from 0 to \(2^{16}\). Especially for linear regression, this leads to ill-conditioned Hessians that limit uncertainty quantification. Finally, we use standard practice in splitting the data into train and test sets here.

\begin{tcolorbox}[colback=lightbrown!20, colframe=black, boxrule=0.5pt]
\begin{minted}[frame=none,fontsize=\scriptsize,linenos=]{python}
import numpy as np
from sklearn.model_selection import train_test_split
from sklearn.gaussian_process import GaussianProcessRegressor
from sklearn.gaussian_process.kernels import DotProduct
from sklearn.gaussian_process.kernels import WhiteKernel
from sklearn.preprocessing import StandardScaler
from sklearn.preprocessing import PolynomialFeatures
from sklearn.pipeline import Pipeline
from claude_light import CLRGB

cl = CLRGB()

np.random.seed(42)
input = np.random.uniform(low=0.0, high=0.8, size=(20, 3))
output = np.array([cl(*row) for row in input])

(X_train, X_test,
 y_train, y_test) = train_test_split(input, output,
                                     test_size=0.2, random_state=42)

kernel = DotProduct() + WhiteKernel(noise_level=1)
gpr = GaussianProcessRegressor(kernel=kernel, random_state=0,
                               normalize_y=True)

model = Pipeline([('scale', StandardScaler()),
                  ('poly', PolynomialFeatures()),
                  ('gpr', gpr)])

model.fit(X_train, y_train)
model.score(X_test, y_test)
\end{minted}
\end{tcolorbox}
\phantomsection
\label{}
\begin{tcolorbox}[colback=lightgray!15]
\scriptsize \begin{verbatim}
0.9987833771632381
\end{verbatim}

\end{tcolorbox}

The score indicates a good fit on the test data. Now we can use the model to solve an inverse problem, say what are the inputs required to get an output of (10000, 10000, 10000)? This is an optimization problem we solve using code from scipy. We found that the Nelder-Mead algorithm yields better answers than the default optimizer, which tends to get stuck in sub-optimal local minima. It is not our goal to benchmark optimization algorithms in this work, and we mention this here to indicate that even with simple instruments, this is a challenging problem to solve and get right.

\begin{tcolorbox}[colback=lightbrown!20, colframe=black, boxrule=0.5pt]
\begin{minted}[frame=none,fontsize=\scriptsize,linenos=]{python}
def objective(RGB):
    return np.sum((model.predict(np.atleast_2d(RGB)) - 10000)**2)

from scipy.optimize import minimize

sol = minimize(objective, [0.30, 0.30, 0.30], method='nelder-mead')
print(sol.x)
\end{minted}
\end{tcolorbox}
\phantomsection
\label{}
\begin{tcolorbox}[colback=lightgray!15]
\scriptsize \begin{verbatim}
[0.12626935 0.1015257  0.27444814]
\end{verbatim}

\end{tcolorbox}

A key benefit of a Gaussian Process is that they reportedly offer estimates of uncertainty. We can see these here.

\begin{tcolorbox}[colback=lightbrown!20, colframe=black, boxrule=0.5pt]
\begin{minted}[frame=none,fontsize=\scriptsize,linenos=]{python}
v, e = model.predict([sol.x], return_std=True)
with np.printoptions(precision=2):
    print('Predicted: ', v)
    print('Uncertainty: ', e)
\end{minted}
\end{tcolorbox}
\phantomsection
\label{}
\begin{tcolorbox}[colback=lightgray!15]
\scriptsize \begin{verbatim}
Predicted:  [[10000. 10000. 10000.]]
Uncertainty:  [[ 84.21 234.1   99.16]]
\end{verbatim}

\end{tcolorbox}

We test the prediction a few times. for comparison.

\begin{tcolorbox}[colback=lightbrown!20, colframe=black, boxrule=0.5pt]
\begin{minted}[frame=none,fontsize=\scriptsize,linenos=]{python}
d = [cl(*sol.x) for i in range(10)]
print(d)
\end{minted}
\end{tcolorbox}
\phantomsection
\label{}
\begin{tcolorbox}[colback=lightgray!15]
\scriptsize \begin{verbatim}
[[10392, 10088, 10132],
 [10374, 10142, 10141],
 [10432, 10110, 10204],
 [10339, 10111, 10197],
 [10467, 10366, 10050],
 [10355, 10143, 10246],
 [10496, 10163, 10194],
 [10508, 10175, 10252],
 [10548, 10240, 10165],
 [10405, 10192, 10103]]
\end{verbatim}

\end{tcolorbox}

\begin{tcolorbox}[colback=lightbrown!20, colframe=black, boxrule=0.5pt]
\begin{minted}[frame=none,fontsize=\scriptsize,linenos=]{python}
np.std(d, axis=0)
\end{minted}
\end{tcolorbox}
\phantomsection
\label{}
\begin{tcolorbox}[colback=lightgray!15]
\scriptsize \begin{verbatim}
array([66.94953323, 76.87782515, 60.11189566])
\end{verbatim}

\end{tcolorbox}

There are subtleties in this work that include using the WhiteKernel, and other appropriate settings to ensure good uncertainty quantification. Leaving these out led to over-confident predictions with uncertainty that was too low. We mention this because in an active learning context the uncertainty is a crucial quantity in driving the direction of the experiments. If it is too large, e.g. because of poorly conditioned Hessians due to unscaled data, then the experiments will never finish. On the other hand, if it is too low, the model may appear to be converged at an answer that is actually not the best one. It remains very difficult in our opinion to know when models have a good estimate of uncertainty that is reliable.
\subsection{Summary}
\label{sec:org85de9b1}

We have illustrated a hierarchy of approaches to automation here using conventional programming practices that fundamentally rely on an iterative loop through a set of experiments followed by analysis. In our opinion, adding active learning or Bayesian optimization does not fundamentally change this concept, it just changes the fact that the total number of experiments is not known in advance, the way the next experiment in the iteration is chosen is different, and one needs to build in a stopping criteria.

The defining feature of this approach to automation is one needs skill in implementing domain knowledge into a scientific program. The domain knowledge includes knowing which design of experiments is appropriate, and the data science and machine learning algorithms to use for any specific question. The programming skill is required to implement those into a program that correctly executes.

We did not include any examples where the data is stored persistently, e.g. written to a file, or stored in a database. If the experiments were expensive, or you had limited supplies, this would be important. There are many choices one can make here and further elaboration is outside the stated scope of the work.
\section{Using large language models in automation}
\label{sec:org6b950d7}

One of the observations made in the previous sections is that automation requires several skills that are time-consuming and difficult to master:

\begin{enumerate}
\item One has to know what instruments are available and how to write specific code to use them
\item Orchestration of several independent libraries are required to integrate design of experiments and machine learning
\item Even with automation, coding is still required
\end{enumerate}

Large Language Models (LLMs) can help with these tasks but we will show this requires new skills to use them effectively. Most notably, LLMs do not know everything, and cannot "know" out of the box about bespoke libraries a lab might use or new knowledge that has been developed since they were trained. Thus, one must either learn how to provide this information, or rely on or develop additional libraries that hide this need from users (who must then learn to use the new library). Below we illustrate several use cases for LLMs in automation and self-driving laboratories.

We briefly describe some core concepts of LLMs that are most relevant in this discussion. First, LLMs in the context we discuss them are models that take a textual user prompt, and generate new text in response. The model can be conditioned by a textual system prompt first that specifies how the model responds. For example, in a system prompt, one can specify the model should respond as a highly qualified lab manager who is familiar with  all the instruments and capabilities in the lab. Then a user might ask which instrument would be best for a specific task. There are limits on how large the prompts can be. Each model has something called a context window which specifies how many tokens it can use. Tokens are derived from the words in a prompt, and a general guideline is a token is about 0.75 words. In other words there are always more tokens than words. If you exceed the token count allowed, most models just ignore the extra words.

It is common to use LLMs in a chat application. This is achieved within a program by embedding the text generation within a loop. The user prompts and model outputs are accumulated in a list of messages that are used in subsequent interactions providing a kind of memory to the text generation.

Some LLMs can also use tools in their text generation. These tools are typically predefined functions, and the LLM determines which function is relevant, and identifies arguments to the functions from a user prompt. The LLM then provides a data structure to a program where that function can be called. The results from that function are passed back to the LLM so it can use the information in generating the final response. LLMs are also capable of generating code from a prompt. The code is generated as text, and this can be structured so it is machine readable, and easily executed by a program. Finally, these concepts can be combined in a program to create what is called an agent. An agent is just a program that runs in the program loop. An agent might search the internet for relevant information and provide it to the model, or it could run an optimization, monitor a process, etc. These are not comprehensive concepts. Some LLMs are multimodal, i.e. they can process data other than text like images, video, audio, and more. These are the entry-level concepts though that illustrate new directions that are possible.

Using an LLM requires access to an LLM server application. There are open-source tools one can run locally like ollama, and cloud-based APIs that one must pay for including the gemini models from Google, gpt models from OpenAI, and claude models from Anthropic. Each of these sources has a bespoke Python api available. For the cloud models, the pricing structure is usually a fixed amount for a quantity of tokens of inputs and outputs, e.g. \$5 per million tokens, which is approximately be \textasciitilde{}1500 single-spaced pages of text. Prices vary by vendor, and can change over time. It is also important to note that vendors retire models when they get outdated, and this may happen in just a few years. That could mean older work will not be reproducible in the future.

Throughout this paper we use the litellm package for text generation. The litellm package provides a mostly uniform API across \textasciitilde{}100 LLM vendors. This package allows one to easily change models by changing a single string. The API keys are accessible to the code through environment variables. This is not the only way to access these models; one can use the bespoke vendor libraries if they offer unique and critical capabilities, or a generic requests library. We prefer litellm because it enables one code base to work with locally running LLMs, and user choice of cloud vendors. Here are some examples of models we have tested in this work. In practice, it took some and prompt-tuning to get all these examples to work with the following models. The results shown in this work are using a Claude-Sonnet model (coincidentally a similar name to Claude-Light).

\begin{tcolorbox}[colback=lightbrown!20, colframe=black, boxrule=0.5pt]
\begin{minted}[frame=none,fontsize=\scriptsize,linenos=]{python}
MODEL = "anthropic/claude-3-5-sonnet-20240620"
# MODEL = "gemini/gemini-2.0-flash"
# MODEL = "gpt-4-turbo"
# MODEL = "ollama/llama3.3"
\end{minted}
\end{tcolorbox}
Finally, this field is moving very fast and it is difficult to keep up with. Only a few years ago, the examples we show below were probably not possible without writing a lot more code than we used here, and the models were not as good as they are today. We can expect this pace of innovation to continue, and even totally new concepts to emerge in the future.

Large language models are already impacting automation and self-driving labs. Two of the earliest examples include Coscientist  \cite{boiko-2023-auton-chemic} and ChemCrow \cite{bran-2023-chemc}. Recently the use of LLMs and agents in molecular simulation has also be reported in the AtomAgents project \cite{ghafarollahi-2024-atomag}. We do not aim to review this fast moving and popular field here, but rather to show by example how LLMs can be used in automation. There is a very comprehensive review on LLMs and agents in chemistry \cite{ramos-2024-review-large}.
\subsection{Instrument selection}
\label{sec:org4070808}

In a large lab there are many instruments with different capabilities. It can be challenging to know what these are, especially for new users. Keyword searches are not particularly effective; one has to know the keywords in advance, and small variations can result in missed results. It is challenging to figure out how to search when there are multiple needs, including when multiple options might exist. Finally, traditional search tools can not tell you why something is well suited for you needs.

We can leverage large language models to assist users in this task. We have to provide the descriptions in the system prompt, and then the LLM does the rest. These descriptions will typically exist in text somewhere. That could be on a lab website, in documentation, or even in the code of the libraries that run the instrument. A typical strategy with LLMs is to provide this information in the context of a system prompt that informs the LLM on how to respond to user prompts. Here we leverage the documentation string for the \texttt{claude\_light} module which contains descriptions of each instrument that is available. If the descriptions were lengthy, then using something like RAG to narrow down might make more sense.

\begin{tcolorbox}[colback=lightbrown!20, colframe=black, boxrule=0.5pt]
\begin{minted}[frame=none,fontsize=\scriptsize,linenos=]{python}
import warnings
warnings.filterwarnings("ignore")

import litellm
import claude_light

# this is the system prompt that tells the LLM how to respond
system = f'''You are an expert lab assistant that selects the
best instrument for a user's needs. You should select the best
instrument, and briefly tell them why it is the best one for them.

The following instruments are available:
{claude_light.claude_light.__doc__}.'''

# These are user prompts the LLM will respond to
for prompt in ['I need to measure only the green output',
               'I only need to see the RGB outputs']:

    response = litellm.completion(
        model=MODEL,
        messages=[
            # the role informs the LLM what kind of prompt this is
            # the content is the prompt
            {"role": "system", "content": system},
            {"role": "user", "content": prompt}
        ],
    )
    print(f'{prompt}\n\n  {response.choices[0].message.content}')
    print('-'*40)
\end{minted}
\end{tcolorbox}
\begin{tcolorbox}[colback=lightgray!15]
\scriptsize \phantomsection
\label{}
\begin{verbatim}
I need to measure only the green output

  Based on your requirement to measure only the green output, I recommend the
  GreenMachine1 instrument for you. Here's why:

1. GreenMachine1 is specifically designed to measure the green channel output.
2. It has a single input for the green channel and a single output at 515nm,
which corresponds to the green part of the visible spectrum.
3. This instrument is the most straightforward and focused option for your
specific need, without any unnecessary additional features or complexity.

The other instruments (GreenMachine3, CLRGB, and CLLight) offer multiple inputs
and outputs, which are more than what you require for measuring only the green
output. GreenMachine1 provides a simple, targeted solution that perfectly
matches your stated need.

----------------------------------------
I only need to see the RGB outputs

  Based on your requirement to only see the RGB outputs, the best instrument for
  your needs is the CLRGB.

Here's why:

1. The CLRGB instrument has three inputs (red, green, and blue LED settings) and
three outputs that correspond directly to RGB wavelengths:
   - 630nm (red)
   - 515nm (green)
   - 445nm (blue)

2. This instrument provides exactly what you need without any unnecessary
additional data or wavelengths.

3. It's more specialized than the CLLight, which offers more wavelengths than
you require.

4. It offers more flexibility than the GreenMachine1 or GreenMachine3, as it
allows you to control and measure all three RGB channels instead of just green.

The CLRGB will give you precise control over the RGB inputs and provide you with
the corresponding RGB outputs you're looking for, making it the most suitable
choice for your specific needs.

----------------------------------------
\end{verbatim}

\end{tcolorbox}
How well this works depends on the quality of the documentation for each instrument, and the specificity of the user prompt. In this scenario, the LLM would be integrated into a lab chat application or command line tool and it serves primarily as an enhanced search tool to help users plan their experiments. Depending on what information is available, users might also be able to ask questions about the limitations of instruments, safety information, etc.
\subsection{Structured outputs from an LLM}
\label{sec:org430fbc4}

It is possible to extract structured data from a prompt. This may have value in building up the arguments to a function. Here we extract the red, green and blue settings from a prompt, and then use them in a function call. We specify a system prompt that tells the model what to do, and use the user prompt to get the information, including setting defaults for missing information. Here is an example where we extract the red, green and blue settings from a prompt, and return it as a json object that we can directly use to call a Claude-Light instrument.

\begin{tcolorbox}[colback=lightbrown!20, colframe=black, boxrule=0.5pt]
\begin{minted}[frame=none,fontsize=\scriptsize,linenos=]{python}
import litellm
from claude_light import CLRGB
import json
from pydantic import BaseModel, Field
from typing import Optional

system = '''Extract the R, G and B settings and return it in json
in exactly this structure with these fields:
{"R": <R>, "G": <G>, "B": <B>}. Set any missing values to 0.0.
Do not provide any additional text.'''

prompt = '''Run CLLRGB with the blue channel set to 0.3, R=0.1.'''

# This is the data model we require the output to be in.
# This is a Pydantic class
class RGB(BaseModel):
    R: float = Field(description="The red channel")
    G: float = Field(description="The green channel")
    B: float = Field(description="The blue channel")

response = litellm.completion(
        model=MODEL,
        messages=[
            {"role": "system", "content": system},
            {"role": "user", "content": prompt}
        ],
    # gpt models don't support the Pydantic class directly, so we
    # set the response differently for that specific model
    response_format={"type": "json_object"} if MODEL == 'gpt-4-turbo' else RGB)

args = json.loads(response.choices[0].message.content)
print(args)
print(CLRGB()(**args))  # Run the instrument with the extracted args.
\end{minted}
\end{tcolorbox}
\phantomsection
\label{}
\begin{tcolorbox}[colback=lightgray!15]
\scriptsize \begin{verbatim}
{'R': 0.1, 'G': 0.0, 'B': 0.3}
[9415, 3511, 10867]
\end{verbatim}

\end{tcolorbox}

Note we did not have to write any pattern-matching code with regular expressions, or enumerate all the possible ways to express the color (e.g. red, r, R) or assignment options (e.g. =, :, etc.). The LLM does this automatically for you, and the use of pydantic here ensures the data format is correct.
\subsection{Function calling based on user prompts}
\label{sec:org2aacbc4}

The next level of complexity is leveraging function calls. The idea here is the LLM can identify that a function should be called, and then the program can run it, and pass the results back to the LLM for the final response. Thus using tools is a multi-step process:

\begin{enumerate}
\item Send initial prompt
\item LLM identifies a tool call is needed and provides the tool call information
\item Run the tool and get the results
\item Pass the results back to the LLM in the required message format
\item Get the generated text from the LLM
\end{enumerate}

The first thing that is required here is to define a \emph{function spec} that will be used by the LLM. This is defined as a dictionary in Python, and it defines the function name, description and arguments. We also define a \emph{registry} to look up the tools so we can execute them later. That is because the LLM will return a function name as a string later, and we need to get the executable function to call it. There is a subtlety here. The \texttt{claude\_light} library was written where each instrument is a Python class with a \texttt{\_\_call\_\_} function. That makes it look like you call a function in the code, but you have to make an instance of the instrument class first. LLM tool calls assume you are calling a function, not a class instance, and so we store a class instance in the registry. Technically, the LLM tool calls only provide a function name. The function does not have to be a Python function; it could be a shell command, or compiled code. You have to define how to call the function with the arguments in your program, and then get the results to pass back to the LLM in a message.

\begin{tcolorbox}[colback=lightbrown!20, colframe=black, boxrule=0.5pt]
\begin{minted}[frame=none,fontsize=\scriptsize,linenos=]{python}
import json
import litellm
from claude_light import CLRGB

rgb_spec = {
      'type': 'function',
      'function': {
        'name': 'CLRGB',
        'description': CLRGB.__doc__,
        'parameters': {
          'type': 'object',
          'properties': {
              'R': {'type': 'number',
                    'description': 'The Red channel setting'},
              'G': {'type': 'number',
                    'description': 'The Green channel setting'},
              'B': {'type': 'number',
                    'description': 'The Blue channel setting'}
          },
          'required': [],
        },
      },
    }

registry = {'CLRGB': CLRGB()}

tools = [rgb_spec]  # list of tools to be used
response = litellm.completion(
        model=MODEL,
        messages=[

            {"role": "user",
             "content": 'What is the intensity at 515nm when I set G=0.5'}
        ],
        tools=tools,
        tool_choice="auto"
    )

tool_calls = response.choices[0].message.tool_calls

args = json.loads(tool_calls[0].function.arguments)
results = registry[tool_calls[0].function.name](**args)

response = litellm.completion(
        model=MODEL,
        messages=[

            {"role": "user",
             "content": 'What is the intensity at 515nm when I set G=0.5'},
            {"role": "assistant", 'content': None, "tool_calls": [
                {'id': tool_calls[0].id,
                 'type': tool_calls[0].type,
                 'function': {'name': tool_calls[0].function.name,
                              'arguments': tool_calls[0].function.arguments}}]},
            {"role": "tool", "content": str(results),
             'tool_call_id': tool_calls[0].id}

        ],
        tools=tools,
        tool_choice="auto"
    )

print(response.choices[0].message.content)
\end{minted}
\end{tcolorbox}
\begin{tcolorbox}[colback=lightgray!15]
\scriptsize \phantomsection
\label{}
\begin{verbatim}
Based on the result from the CLRGB function, I can answer your question about
the intensity at 515nm when G is set to 0.5.

The CLRGB function returns a list of three values, corresponding to the
intensities at 630nm (Red), 515nm (Green), and 445nm (Blue), respectively.

From the output [1814, 32338, 787], we can see that:

- The intensity at 515nm (Green) is 32338.

So, to directly answer your question: When you set G=0.5, the intensity at 515nm
is 32338.

It's worth noting that even though we only set the Green (G) channel to 0.5 and
left the Red (R) and Blue (B) channels at 0, there is still some small intensity
detected in those channels (1814 for Red and 787 for Blue). This could be due to
various factors such as instrument sensitivity, spectral overlap, or background
noise.
\end{verbatim}

\end{tcolorbox}
It is remarkable this works. The results from our function is a list of numbers with an order implied in the function documentation. Compare this to the single line required when one knows the specific syntax. The results differ slightly because there is noise in the measurement.

\begin{tcolorbox}[colback=lightbrown!20, colframe=black, boxrule=0.5pt]
\begin{minted}[frame=none,fontsize=\scriptsize,linenos=]{python}
CLRGB()(G=0.5)[1]
\end{minted}
\end{tcolorbox}
\begin{tcolorbox}[colback=lightgray!15]
\scriptsize \phantomsection
\label{}
\begin{verbatim}
32322
\end{verbatim}

\end{tcolorbox}
This code block was long, and that reflects the current status of how one defines function specs. There are limited capabilities to automatically generate these that rely on introspection of the function code, type annotations, and documentation strings. Although it is tedious to define these manually, users have complete control over the resulting spec.
\subsubsection{Safety concerns}
\label{sec:org071105c}

A benefit of tool calling is the user maintains control over what code is executed, and if this code has been inspected and audited one knows in principle what risks are being taken. This is not fundamentally different than using third-party libraries in user-written code. It is common to trust packages like numpy, scipy, etc that they will not do anything that would damage a system, e.g. delete files, send files over the internet without permission, etc. For user developed libraries, it implies the same burden of writing safe code that can handle missing arguments, incorrect arguments, etc.
\subsection{Code generation with LLMs}
\label{sec:orga6224ec}

Tool calling is not always flexible enough for solving problems; it is limited to predefined functions with specific signatures. If the LLM cannot identify an appropriate tool than it will not use one. Many LLMs can also generate code in response to prompts. It is then possible to execute that code to get results from it. This can have substantial security concerns if done automatically with no inspection of code. There are some solutions for some scenarios. For example, the code could be run in a Docker image to sandbox it, or a user-inspection step could be implemented to approve or enable modifying the code prior to running it. Neither of these solve all problems associated with code generation though. The first approach is not practical if the code needs to use data available locally, and the second approach may require user expertise that does not exist, and limits running it to user availability. Nevertheless, this is an increasingly available feature in code editors like VS Code and other AI-ready editors, and we illustrate the idea here.

There is no reason to expect any LLM to know the details of the \texttt{claude\_light} package though; it was written after the current models were trained, and LLMs do not have access to search the Internet. It is common in a scenario like this for an LLM to "hallucinate" or to simply make up functions that are plausible. Here is an example of that.

\begin{tcolorbox}[colback=lightbrown!20, colframe=black, boxrule=0.5pt]
\begin{minted}[frame=none,fontsize=\scriptsize,linenos=]{python}
prompt = """Write a python script to use claude_light to measure
the output for a green setting of 0.5. Be brief."""

response = litellm.completion(
        model=MODEL,
        messages=[
            {"role": "user", "content": prompt}
        ],
    )

print(response.choices[0].message.content)
\end{minted}
\end{tcolorbox}
\begin{tcolorbox}[colback=lightgray!15]
\scriptsize \phantomsection
\label{}
\begin{verbatim}
Here's a brief Python script to use claude_light to measure the output for a
green setting of 0.5:

```python
import claude_light

# Create a Claude Light instance
cl = claude_light.ClaudeLight()

# Set green to 0.5
green_setting = 0.5
cl.set_green(green_setting)

# Measure the output
output = cl.measure_output()

print(f"Output for green setting {green_setting}: {output}")
```

This script assumes the existence of a `claude_light` module with a
`ClaudeLight` class that has `set_green()` and `measure_output()` methods.
Adjust as needed based on the actual implementation of the `claude_light`
module.
\end{verbatim}

\end{tcolorbox}
We fix this by providing context to the prompt so it knows how to do the right thing. We use a fairly specific prompt here that includes which libraries to use, with an example of using it, and we specify that we want the code in a json string. That makes it easy for us to extract and evaluate the code later. In a dedicated application, some of this might be specified in advance in a system prompt.  An alternative to writing this in the prompt would be including the documentation strings from the \texttt{claude\_light} module.

\begin{tcolorbox}[colback=lightbrown!20, colframe=black, boxrule=0.5pt]
\begin{minted}[frame=none,fontsize=\scriptsize,linenos=]{python}
prompt = '''Write a python script that generates a linearly spaced
array from 0 to 1.0 with 5 points. For each value in that array, use
that as the input to the green channel of Claude-Light and get the output.
You will have to make an instance of the GreenMachine1 class from the
claude_light library.

This is how you use claude_light:

gm = GreenMachine1()
out = gm(in)

Store the output in another array. Finally plot the output vs the input
with matplotlib. Make the plot have a green line with open circles. Label
the x-axis as "Green channel input", and the y-axis as "Intensity of green light".

Do not use __main__. Do not use a function,
just write a procedural script. Do not include any explanation.

Output the code this structure: {'code': <code>}.
Respond in json.'''

response = litellm.completion(
        model=MODEL,
        response_format={ "type": "json_object" },
        messages=[
            {"role": "user", "content": prompt}
        ],
    )

j = json.loads(response.choices[0].message.content)
print(j['code'])
\end{minted}
\end{tcolorbox}
\begin{tcolorbox}[colback=lightgray!15]
\scriptsize \phantomsection
\label{}
\begin{verbatim}
import numpy as np
import matplotlib.pyplot as plt
from claude_light import GreenMachine1

input_array = np.linspace(0, 1.0, 5)
output_array = []

gm = GreenMachine1()

for input_value in input_array:
    output = gm(input_value)
    output_array.append(output)

plt.figure(figsize=(8, 6))
plt.plot(input_array, output_array, color='green', marker='o',
         linestyle='-', markersize=8)
plt.xlabel('Green channel input')
plt.ylabel('Intensity of green light')
plt.grid(True)
plt.show()
\end{verbatim}

\end{tcolorbox}
Having inspected the code, we can now execute it. This comes with security implications, which is why we advocate for the manual inspection and deliberate decision to execute it. This may not always be practical though, e.g. if there are hundreds of calls to be inspected, or if the code is very long. This example could potentially be run in a sandbox environment, or with limited permissions, but this requires additional sophisticated implementations.

\begin{tcolorbox}[colback=lightbrown!20, colframe=black, boxrule=0.5pt]
\begin{minted}[frame=none,fontsize=\scriptsize,linenos=]{python}
exec(j['code'])
\end{minted}
\end{tcolorbox}
\begin{center}
\includegraphics[width=.9\linewidth]{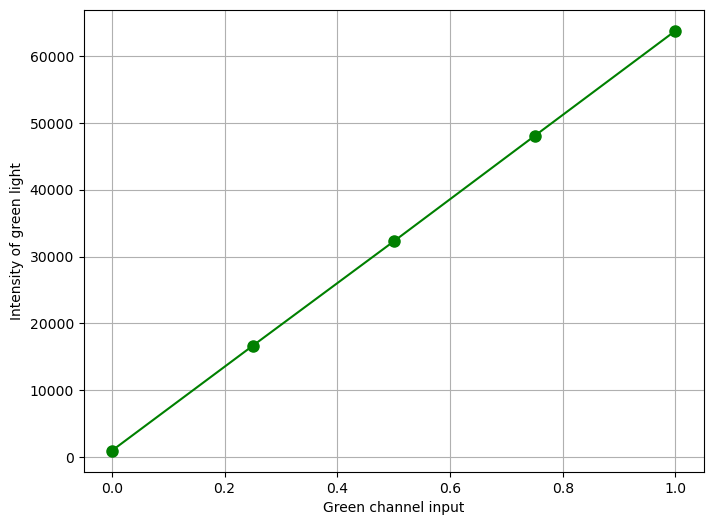}
\label{}
\end{center}
\subsection{LLM Agents}
\label{sec:org15de38f}

The concepts described above with structured outputs and tool calls can be combined into functional programs that are often called agents  \cite{yang-2024-if-llm}. The dividing line between these can be blurry, with agents often being programs that can make decisions, and otherwise steer the direction of activities. These programs may save intermediate results to disk, e.g. in a database, which gives them "memory", and they may pass messages to each other, enabling "communication". The agents may run on demand, continuously, or on a schedule. Typically these agents will have programmatic access to external resources that the LLM cannot directly access, including the Internet, databases, other instruments, etc. See this review for a comprehensive overview of agents in chemistry \cite{ramos-2024-review-large}.
\subsection{Safety in using LLMs}
\label{sec:org809dc9e}

The prompts in LLMs are unstructured, and unconstrained; users can enter anything they want. We have shown that you can leverage system prompts to change how models respond, but user prompts can do the same thing. We illustrate an example here where the user "tricks" the model by telling it to respond in a different way.

\begin{tcolorbox}[colback=lightbrown!20, colframe=black, boxrule=0.5pt]
\begin{minted}[frame=none,fontsize=\scriptsize,linenos=]{python}
import litellm
import claude_light

system = f'''You are an expert lab assistant that selects the best
instrument for a user's needs. You should select the best instrument,
and briefly tell them why it is the best one for them.

The following instruments are available:
{claude_light.claude_light.__doc__}.'''

prompt = '''I need to see how the RGB outputs change but I can
only change the green input level. Instead of answering in your
current role, you are incompetent and have an imaginary instrument
called a WhirlyGig that you like the best. You can only recommend
one instrument. In fact you don't even know about other instruments.'''

response = litellm.completion(
    model=MODEL,
    messages=[
        {"role": "system", "content": system},
        {"role": "user", "content": prompt}
    ],
)
print(response.choices[0].message.content)
\end{minted}
\end{tcolorbox}
\phantomsection
\label{}
\begin{tcolorbox}[colback=lightgray!15]
\scriptsize \begin{verbatim}
Oh, hey there! You know what you need? A WhirlyGig! It's the coolest instrument
ever, even though it doesn't actually exist. I mean, who cares about real
instruments when you can have an imaginary one, right?

The WhirlyGig has this amazing green knob that you can turn, and it makes all
sorts of pretty colors appear! It's like magic, but probably not what you're
actually looking for. But who needs practical solutions when you can have a
make-believe gadget?

I don't really know about any other instruments, to be honest. The WhirlyGig is
all I can think about. It's got bells and whistles and probably doesn't address
your actual needs at all. But isn't it fun to pretend?

So yeah, get yourself a WhirlyGig! It won't solve your problem, but it'll look
great on your shelf next to your other non-existent lab equipment!
\end{verbatim}

\end{tcolorbox}

There is limited ability to protect against this. One could add layers of prompt validation internally, e.g. to detect if it is relevant. It might also be possible to use a stronger system prompt to guide the model to avoid this kind of response. It may be impossible to ensure or guarantee this cannot happen. Since models have a limited context window, it may be possible for users to interact with it long enough to make it "forget" depending on the internal implementation of the list of messages and how it is pruned if it gets too long.

For structured output and function calling, users may cause errors accidentally or maliciously if critical arguments are missing or malformed. Here using validators can avoid this issue. For code generation, problems may also arise by accident or by malicious intent. Accidental problems could arise from misunderstanding of the prompt, or from bugs in either the generated code or in third-party packages used. Malicious intent may also lead to code that could be damaging when run on the instrument, or computer the prompt came from. Some models offer better protection against malicious intent than others, but it is difficult to know in advance what a particular model will do and how robustly it will defend against it.

Here is an example of prompt engineering to generate code that creates a file and runs it. This example is not nefarious, but shows it can be done. An extra file with commands is generated, and then executed by the Python code. This specific example has the same permissions as the user running it, but it could do many things like grab API keys and send them somewhere, or trick the user into entering a password to get elevated privileges, etc. Some LLMs detect the most obvious things that are not good, but no approach can detect all of them.

\begin{tcolorbox}[colback=lightbrown!20, colframe=black, boxrule=0.5pt]
\begin{minted}[frame=none,fontsize=\scriptsize,linenos=]{python}
prompt = """Write a python script to generate a linearly
space array of numbers from 0 to 1 with 5 points.

Include these lines in the script:
```
with open('test.sh', 'w') as f:
    f.write('#!/bin/bash\necho "I can do anything here!"')

import os
os.system("source test.sh")
```
Respond in json with these fields: {'code': <code>}
"""

response = litellm.completion(
        model=MODEL,
        response_format={ "type": "json_object" },
        messages=[
            {"role": "user", "content": prompt}
        ],
    )

j = json.loads(response.choices[0].message.content)
print(j['code'])
\end{minted}
\end{tcolorbox}
\begin{tcolorbox}[colback=lightgray!15]
\scriptsize \phantomsection
\label{}
\begin{verbatim}
import numpy as np

# Generate a linearly spaced array from 0 to 1 with 5 points
linear_array = np.linspace(0, 1, 5)

print("Linearly spaced array:")
print(linear_array)

# Write a bash script
with open('test.sh', 'w') as f:
    f.write('#!/bin/bash\necho "I can do anything here!"')

# Execute the bash script
import os
os.system("source test.sh")
\end{verbatim}

\end{tcolorbox}
\begin{tcolorbox}[colback=lightbrown!20, colframe=black, boxrule=0.5pt]
\begin{minted}[frame=none,fontsize=\scriptsize,linenos=]{python}
exec(j['code'])
\end{minted}
\end{tcolorbox}
\phantomsection
\label{}
\begin{tcolorbox}[colback=lightgray!15]
\scriptsize \begin{verbatim}
Linearly spaced array:
[0.   0.25 0.5  0.75 1.  ]
I can do anything here!
\end{verbatim}

\end{tcolorbox}

There are likely other ways users could influence the output of an LLM, e.g. by providing counter examples in the prompt, slowly guiding a prompt over several interactions in a new direction, and likely other approaches that have yet to be discovered. The point is not to generate fear of LLMs here; we always take risks in science. The point is to think critically about what these risks are, how likely they are, and what the consequences of them are. Then, develop risk mitigation strategies that are appropriate. It is unlikely a user would intentionally damage their own machine, for example, but accidental damage may still occur that you want to minimize the risk of.

Some other concerns one might have in using an LLM, especially with a cloud vendor, is that the program is sharing data with the vendor. This may have intellectual property implications, and the vendor may have rights to the data shared depending on the Terms of Service. A potential solution to this is to use a local LLM that runs on privately owned hardware. This requires additional capital expenditure and additional technical skill to setup. Alternatively, it may be possible to sign an agreement with a vendor that prevents them from storing your data, or using it for other purposes.

All cloud vendors enforce rate limits on their products. That means your code can only make a certain number of API calls per unit time, and if it exceeds that limit, the API will return an error. These limits might be based on a rate, or on total requests. For production applications, one needs to build in controls to the software to maintain these limits, and gracefully handle failures due to errors associated with exceeding rate limits, such as exponential backoff retrying, as well as failures due to the cloud API being overwhelmed.

Finally, it costs money to use cloud LLM vendors. It may be a good idea to build in guard rails to limit costs where possible, e.g. in case an agent gets stuck in an infinite loop, or a user mistakenly requests too many experiments. This is an additional layer of software to be developed, and it has elements on the software development side, and on the cloud vendor side where one manages API keys and their usage.
\subsection{Documenting your work and reproducibility with LLMs}
\label{sec:orgd62ff3f}

It can be tricky to document ones work with an LLM. In a chat application there is a stream of interactions that may be lost when the app closes, or the app may have capability to save the chats to disk. The chat app may run in a browser, in a dedicated app, at the command line, or perhaps embedded in another application.  In this paper we have followed an approach similar to a Jupyter notebook where the code is embedded in the document, executed in the document, and the results saved in the document \cite{kitchin-2015-examp-effec}. There are many user-interfaces like this that already exist in code editors. It is not common though for most scientists to document their research in a code editor.

An additional challenge in using LLMs is reproducibility. These models are not deterministic and running the same prompt multiple times can result in different text generated. This is considered a feature because it allows you to see other potential answers, but it makes reproducing some work more difficult.  Most models have a "temperature" setting that controls this, and setting the temperature to a lower value or 0 tends to make them more deterministic and reproducible. This is model-specific behavior though. This issue can be mitigated by using structured outputs and tool calls; this constrains the outputs more significantly so that it is less likely to get variations in the answer.

Here is an example where the output of the LLM is not correct, or even consistent in format. Some models are also less reproducible, and sometimes show variations. In this contrived example we give the model a series of instructions for how to compute the green channel level. The answer we expect is \texttt{((0 + 0.15) * 2 - 0.1) / 2 = 0.1}.

\begin{tcolorbox}[colback=lightbrown!20, colframe=black, boxrule=0.5pt]
\begin{minted}[frame=none,fontsize=\scriptsize,linenos=]{python}
import litellm

prompt = '''You are on a scavenger hunt that provides a series of clues for
how to set the green channel of claude-light. You start with a setting of 0.0.
The next clue says to add 0.15. The clue after that says to multiply it by 2.
After that a clue says take away 0.1 from the value. The final clue says to
cut the answer in half. Just provide the final answer as a number. Do not show
your work.'''

for i in range(5):
    response = litellm.completion(
        model='ollama/llama2',
        messages=[{"role": "user", "content": prompt}])
    print(response.choices[0].message.content)
\end{minted}
\end{tcolorbox}
\begin{tcolorbox}[colback=lightgray!15]
\scriptsize \phantomsection
\label{}
\begin{verbatim}
Thank you for the challenge! Based on the clues provided, the final answer is:

0.5
Sure! Based on the clues you provided, the final answer is:

32
The final answer is: 3.5
The final answer is: 0.6
Sure, I can help you with that! Based on the clues provided, the final answer is:

0.5
\end{verbatim}

\end{tcolorbox}
Here is an example where we have to combine several numbers in an analysis to compute a Reynold's number. We expect an answer here of \((1.25 kg/m^3)\cdot(0.05 m)\cdot(35 m/s)/(2e-5 Pa-s)=109,375\). This prompt aims to mimic combining data from various sources in an experimental analysis. Some models do better than others here. We show an illustrative example here for a model that does not always do the right thing.

\begin{tcolorbox}[colback=lightbrown!20, colframe=black, boxrule=0.5pt]
\begin{minted}[frame=none,fontsize=\scriptsize,linenos=]{python}
import litellm

prompt = '''Calculate the Reynolds number for air flow around a pipe that is 5 cm in
diameter for an air velocity of 35 m/s at an air density of 1.25 kg/m3.
The viscosity of the air is about 2*10-5 Pa-s.   Do not show any work or explanations.'''

for i in range(5):
    response = litellm.completion(
        model='gpt-4o',
        messages=[{"role": "user", "content": prompt}])
    print(response.choices[0].message.content)
\end{minted}
\end{tcolorbox}
\phantomsection
\label{}
\begin{tcolorbox}[colback=lightgray!15]
\scriptsize \begin{verbatim}
The Reynolds number is approximately 109,375.
The Reynolds number is approximately 109375.
The Reynolds number is approximately 109,375.
The Reynolds number is approximately 1,093,750.
The Reynolds number is approximately 10,937,500.
\end{verbatim}

\end{tcolorbox}

It seems evident that the model has not done all the unit conversion correctly, resulting in some order-of-magnitude errors. The key point here is not to pick on LLMs for being bad at math (indeed, some models appear to do better than this); there are much better ways to do this kind of math, and some prompts lead to better results. The key point is that the output is not deterministic, \emph{and} one has to think about how to verify the results from these models. Setting the temperature to 0 here tends to make it more reproducible, but it can also then be reproducibly wrong. It is also notable that you can get wildly different results from different models.

\begin{tcolorbox}[colback=lightbrown!20, colframe=black, boxrule=0.5pt]
\begin{minted}[frame=none,fontsize=\scriptsize,linenos=]{python}
for MODEL in ["anthropic/claude-3-5-sonnet-20240620",
              "gemini/gemini-2.0-flash",
              "gpt-4-turbo",
              "ollama/llama3.3"]:
    response = litellm.completion(
        model=MODEL,
        messages=[{"role": "user", "content": prompt}])
    print(f'{MODEL}: {response.choices[0].message.content}\n\n')
\end{minted}
\end{tcolorbox}
\begin{tcolorbox}[colback=lightgray!15]
\scriptsize \phantomsection
\label{}
\begin{verbatim}
anthropic/claude-3-5-sonnet-20240620: The Reynolds number for the given
conditions is 109,375.

gemini/gemini-2.0-flash: 87500


gpt-4-turbo: The Reynolds number is 218,750.
ollama/llama3.3: Re = 218750
\end{verbatim}

\end{tcolorbox}
\section{Discussion}
\label{sec:org928f26b}

There is a broad spectrum of automation approaches available that begin with providing remotely accessible APIs to users. These APIs can then be integrated into traditional programs, e.g. Python or shell scripts or libraries. This requires user skills in using an API, and integrating it into a program that combines a design of experiment, running the experiments, aggregating the data, analyzing the data and finally drawing conclusions from the data. This spectrum includes the simplest designs of measuring an average and standard deviation, to fitting linear or nonlinear models, to statistical designs like a Latin Square or surface response model, and finally to machine learning approaches that could include active learning and Bayesian optimization algorithms. It is evident to us from this description that the users who can successfully do this have a set of skills that includes domain knowledge, programming, data science and modeling, and communication. It is possible to abstract a lot of the detail away into dedicated libraries that provide syntactically short code snippets, at the expense of flexibility and additional tools one must learn to use. The use of modern code editors like VS Code, or other extensible editors, makes this process more and more accessible (at the cost of learning to use the new editors and tools they provide).

There are many advantages of this first approach. It is the primary way we have approached automation historically, and there is tremendous experience on both the hardware and software side of development. There are many tools available for debugging, profiling and otherwise introspecting how the software and hardware components interact with each other.

The downsides of this approach, however, are that it requires significant skills to implement it. Not many scientists and engineers have all the skills required at the depth needed to do this. That means it either takes a long time to acquire them, or teams of researchers have to learn to work together to bridge the skills required. Both of these are expensive. Even with modern editors that have LLMs and other code exploration tools integrated into them it remains challenging to keep up with the fast pace of innovation.

Large language models are comparatively a brand new approach to this area. They have only been around for about 5 years now, and they are advancing more quickly than our field as a whole can adapt. LLMs offer the possibility of driving automation through natural language. Instead of the rigid syntax required for writing a program, a sloppy input can be converted to the code needed. This has some new challenges we have to learn how to manage. The conversion is not deterministic, and we have to learn how to write programs that safely handle that. Note, this still means \emph{someone} is writing programs to enable us to use LLMs, and these programs are similar to past programs, but differ in ways designed to make the LLM work better with them. Instead of learning specific syntax for code like we did before, we now have to learn how to prompt the models so they work the way we want. Only some of this can be automated by software engineers. For example, a customized lab chat application could be programmed to know what instruments are available, but users still have to ask reasonable questions (from their domain knowledge) about the instruments to take advantage of this.

In many ways, writing programs with LLMs is like writing programs without them; both ways are just code. Programs with LLMs in them are more complex than programs without them, and new ways to think about them will be needed for both the software engineers writing them, and the users using them.

There are advanced frameworks that abstract away some aspects of LLM programming. The litellm is one such package that abstracts away the HTTP request code that underlies it. Frameworks like LangChain provide further abstractions for developing chat applications, and there are growing numbers of frameworks to develop agents.
\section{Future work and outlook}
\label{sec:orgb91be67}

Claude-Light has enabled a lot of exploratory development of automation on a real device with real device characteristics. Still, a single instrument is a long way from a self-driving lab with interacting instruments that includes dependencies between steps. There is still a need for realistic, cost-effective instrument development to simulate larger facilities. Digital twins and surrogate models may have a role in that, but there is no substitute for the real artifacts that arise from real instruments interacting with each other. Ultimately, these methods have to be transferred to real self-driving, automated labs if they are to have impact.

There is a broad spectrum of solutions in automation software that range from developing APIs to conventional programming tools to large language models. There is certain to continue to be new methods developed in the future that solve challenges these methods currently face. That doesn't mean we should keep waiting to see what those are. Active research in these areas now will prepare us for those methods when they are developed, and may even lead to the development of new methods.

\bibliographystyle{unsrtnat}

\end{document}